\documentclass[aps,pra, twocolumn, amsmath, amssymb, showpacs, superscriptaddress]{revtex4-2}

\usepackage{xcolor}

\usepackage{graphicx}
\usepackage{dcolumn}
\usepackage{bm}

\begin{document}

\preprint{APS/123-QED}

\title{$T^{-3}$-shift in a short-baseline atomic interferometer-gravimeter}

\author{D. N. Kapusta}
 \email{dmitriikapusta@mail.ru}
 \affiliation{Institute of Laser Physics SB RAS, 630090, Novosibirsk, Russia}
\author{A. E. Bonert}
\affiliation{Institute of Laser Physics SB RAS, 630090, Novosibirsk, Russia}
\author{A. N. Goncharov}
\affiliation{Institute of Laser Physics SB RAS, 630090, Novosibirsk, Russia}
\affiliation{Novosibirsk State Technical University, 630073, Novosibirsk, Russia}
\author{V. I. Yudin}
\affiliation{Institute of Laser Physics SB RAS, 630090, Novosibirsk, Russia}
\affiliation{Novosibirsk State Technical University, 630073, Novosibirsk, Russia}
\affiliation{Novosibirsk State University, 630090, Novosibirsk, Russia}
\author{K. N. Adamov}
\affiliation{Institute of Laser Physics SB RAS, 630090, Novosibirsk, Russia}
\affiliation{Novosibirsk State University, 630090, Novosibirsk, Russia}
\author{A. V. Taichenachev}
\affiliation{Institute of Laser Physics SB RAS, 630090, Novosibirsk, Russia}
\affiliation{Novosibirsk State University, 630090, Novosibirsk, Russia}
\author{M. Yu. Basalaev}
\affiliation{Institute of Laser Physics SB RAS, 630090, Novosibirsk, Russia}
\affiliation{Novosibirsk State Technical University, 630073, Novosibirsk, Russia}
\affiliation{Novosibirsk State University, 630090, Novosibirsk, Russia}
\author{M. D. Radchenko}
\affiliation{Institute of Laser Physics SB RAS, 630090, Novosibirsk, Russia}
\affiliation{Novosibirsk State Technical University, 630073, Novosibirsk, Russia}
\affiliation{Novosibirsk State University, 630090, Novosibirsk, Russia}
\author{O. N. Prudnikov}
\affiliation{Institute of Laser Physics SB RAS, 630090, Novosibirsk, Russia}
\affiliation{Novosibirsk State University, 630090, Novosibirsk, Russia}

\date{\today}

\begin{abstract}
This paper presents the first experimental observation and investigation of a lineshape-asymmetry-caused shift (LACS) in a short-baseline atomic interferometer-gravimeter. It is shown that this shift scales inversely with the cube of the free evolution time, $\propto T^{-3}$, and can lead to a noticeable systematic error in the measured value of the gravitational acceleration~g at the level of 0.1–1 mGal ($T\approx$~milliseconds). The obtained results are in good agreement with our previous theoretical studies and highlight the importance of accounting for LACS in high-precision absolute measurements of g in compact atomic gravimeters.
\end{abstract}

\maketitle

The realization of an atomic interferometer gravimeter based on ultracold atoms in 1991 \cite{Kasevich_1991} marked
the beginning of high-precision inertial quantum sensors. Significant progress has since been achieved in absolute quantum gravimeters (QG) designed for measuring the gravitational acceleration g. Due to their high sensitivity, long-term stability, and accuracy, QGs are currently actively applied in fundamental physics \cite{Rosi_2014,Asenbaum_2020,Zhou_2022}, navigation \cite{Narducci_2022}, geophysics \cite{Stray_2022}, etc.

Miniaturization is one of the most promising directions for QG development. This is facilitated by single-beam cooling methods using diffraction gratings \cite{McGehee_2021,Lee_2022}, progress in compact laser systems \cite{Schkolnik_2016}, including those based on photonic integrated circuits \cite{Lee_2022}, advances in small-scale vacuum chambers \cite{Little_2021}, etc. A distinctive feature of compact atomic interferometer-gravimeters is the short interferometric sequence, typically not exceeding a few tens of milliseconds. This ensures a wide dynamic range and high measurement rates, as well as reduces sensitivity to various systematic shifts and noise.

In contrast, reducing the flight time of atoms in the interferometer decreases sensitivity and long-term stability of the measurements, which is resulting from the quadratic dependence of the phase shift on the interference time. However, in compact QGs, atoms do not drift far from the center of the magneto-optical trap (MOT) during free fall, which allows them to be effectively recaptured and reused in subsequent measurement cycles. Therefore, implementing atom recapture together with high repetition rates (tens to hundreds of hertz) remarkably enhances the signal-to-noise ratio of the interferometric signal and, consequently, improves the metrological performance of QG to levels required for a wide range of practical applications. For example, in \cite{McGuinness_2012}, a short-baseline atomic gravimeter with a maximum sensitivity better than 1~mGal/Hz$^{1/2}$ and a measurement rate of up to 330~Hz is demonstrated. An additional advantage of compact QGs is the reduced requirements for the cold-atom source and the laser system due to efficient atom recapture, which significantly simplifies the overall QG design. Thus, compact atomic gravimeters can be considered a distinct class of gravitational sensors with a number of specific advantages and applications, and warrant further detailed study.

In this context, the study of systematic effects limiting the accuracy of absolute measurements of g in compact atomic interferometers is of both scientific and practical importance. For example, in our recent theoretical work \cite{Yudin_2025}, we investigated a lineshape-asymmetry-caused shift (LACS) in atomic interferometers, which had not previously been discussed in the literature. It was shown that this shift exhibits an inverse cubic dependence $\propto T^{-3}$ on the free-evolution time $T$ and can contribute significantly to the systematic error of a short-baseline atomic interferometers. In this work, we have experimentally observed and studied this LACS shift in an atomic interferometer-gravimeter for the first time. The obtained results are in good agreement with the theoretical predictions in \cite{Yudin_2025} and highlight the importance of accounting for the LACS effect in measurements of the absolute value of g in compact atomic gravimeters.

The experimental setup employed in this study is described in detail in \cite{Kapusta_2025,Bonert_2024}. Each experimental cycle, lasting approximately 100~ms, began with the trapping and cooling of $^{87}\!$Rb atoms to a temperature of about 180~$\mu$K in a conventional MOT. Further sub-Doppler cooling was performed in an optical molasses to a temperature of about 4~$\mu$K after switching off the MOT magnetic field and compensating the residual magnetic fields.

Preparation of the atoms in the magnetically insensitive quantum state $|F=1, m_F=0\rangle$ was performed using optical pumping \cite{Atoneche_2017} according to the following scheme. After the cooling stage, the atoms were pre-pumped to the lower hyperfine level of the ground state $F=1$ by applying a 70~$\mu$s light pulse tuned to the optical transition $|F=2\rangle \leftrightarrow |F'=2\rangle$. A uniform vertical magnetic field of 0.5~G was then applied to define the quantization axis for the Zeeman sublevels. Optical pumping into the state $|F=1, m_F=0\rangle$ was subsequently performed using a 25~$\mu$s pulse of linearly polarized light resonant with the transition $|F=1\rangle \leftrightarrow |F'=0\rangle$. The optical pumping mechanism relied on the depletion of the populations of the magnetic sublevels $|F=1, m_F\pm1\rangle$ and the accumulation of atoms in the state $|F=1, m_F=0\rangle$ via light-induced $\sigma^\pm$ transitions and spontaneous decay. To facilitate these $\sigma^\pm$ transitions, the propagation direction of the pumping beam was aligned parallel to the magnetic-field vector. The optical pumping efficiency exceeded 95\,$\%$, with negligible heating of the atomic cloud (approximately 1~$\mu$K). The intensity of the laser beams used at this stage did not exceed 0.5~mW/cm$^2$.

\begin{figure}
    \centering
    \includegraphics[width=1.00\linewidth]{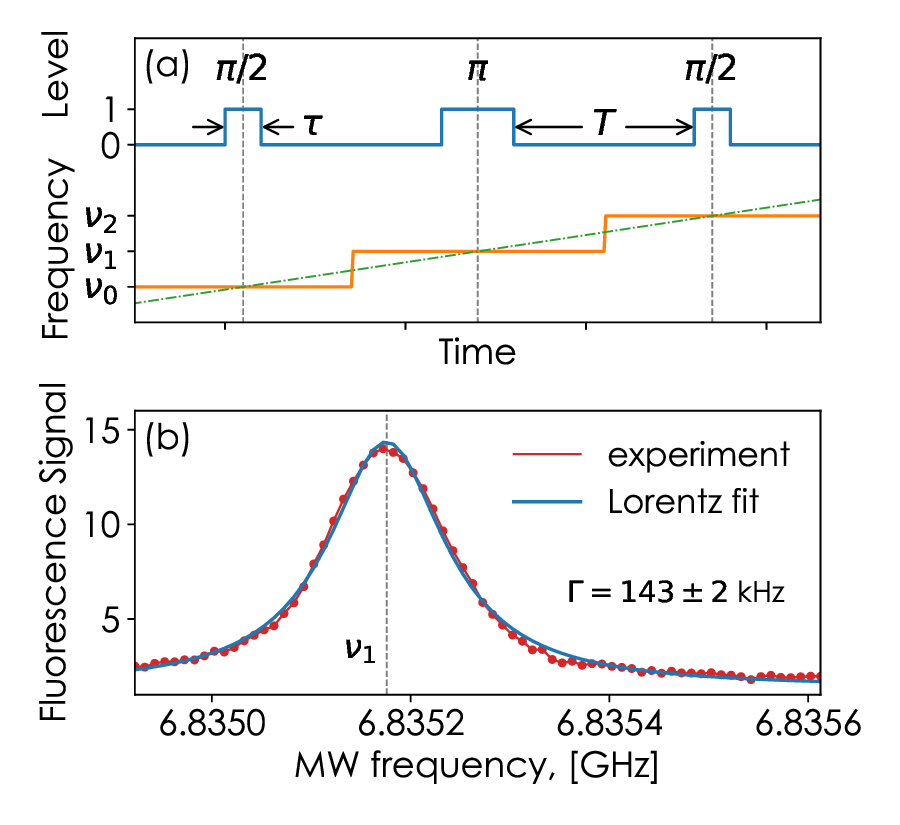}
    \caption{(Color online) Chirping of the Raman beam frequency: (a) interferometric sequence (blue curve) and frequency chirp scheme (orange curve); (b) typical resonance corresponding to the two-photon transition $|F=1, m_F=0\rangle \rightarrow |F=2, m_F=0\rangle$.}
    \label{fig:raman_seq}
\end{figure}

In the next step, the atomic interference stage was implemented using two-photon stimulated Raman transitions $|F=1, m_F=0\rangle \leftrightarrow |F=2, m_F=0\rangle$. An atom interferometer scheme with a $\pi/2-\pi-\pi/2$ pulse sequence was employed to perform coherent splitting, inversion, and recombination of atomic wave packets \cite{Kasevich_1992}. The stimulated transitions were driven by counter-propagating Raman beams aligned parallel to the vector $\textbf{g}$, with mutually orthogonal linear polarizations. The beams had a diameter of 1~cm (FWHM) and a power of 50~mW. The Raman beam radiation was generated using a phase electro-optic modulator (EOM) and included two frequency components red-detuned from the $|F=1,2\rangle \leftrightarrow |F'=1\rangle$ transitions (D2 line) by 1.2~GHz. The frequency of the Raman beams was tuned by varying the microwave signal frequency applied to the EOM input near the ground-state hyperfine splitting frequency $\nu_{hfs}\approx$~6.835~GHz. To compensate for the Doppler shift and form the interference pattern, a chirping technique was applied to the differential frequency of the Raman beams (Fig.~\ref{fig:raman_seq}(a)). The chirp was performed stepwise in the intervals between the Raman pulses by tuning the microwave frequency. The frequency for each pulse was selected in accordance with $\nu_n = \nu_1 + (n-1) \alpha (T+1.5\tau) + \delta_D$, where $n=1,2,3$ is the pulse index, $\nu_1$ is the frequency of the first pulse, $\alpha$ is the chirp rate, $\tau=20~\mu$s is the $\pi/2$-pulse duration, and $\delta_D$ is the controlled detuning used for LACS shift registration (see below). The frequency $\nu_1$ corresponded to the resonant frequency of the two-photon transition $|F=1, m_F=0\rangle \rightarrow |F=2, m_F=0\rangle$, including all relevant shifts. To determine $\nu_1$, this resonance was preliminarily detected by scanning the Raman beam frequency. For this purpose, after optical pumping, the atoms were exposed only to the first Raman pulse of the interferometric sequence and transferred to the $|F=2, m_F=0\rangle$ state. Detection was performed via the fluorescence signal using an approach described below. The resulting frequency dependence of the fluorescence signal (Fig.~\ref{fig:raman_seq}(b)) was fitted with a Lorentzian function to measure $\nu_1$. The discrepancy between $\nu_1$ and $\nu_{hfs}$ is primarily due to Doppler shift (free-fall time of the atoms $\approx$ 18.7~ms) and residual Stark shift.

\begin{figure*}
    \centering
    \includegraphics[width=0.95\textwidth]{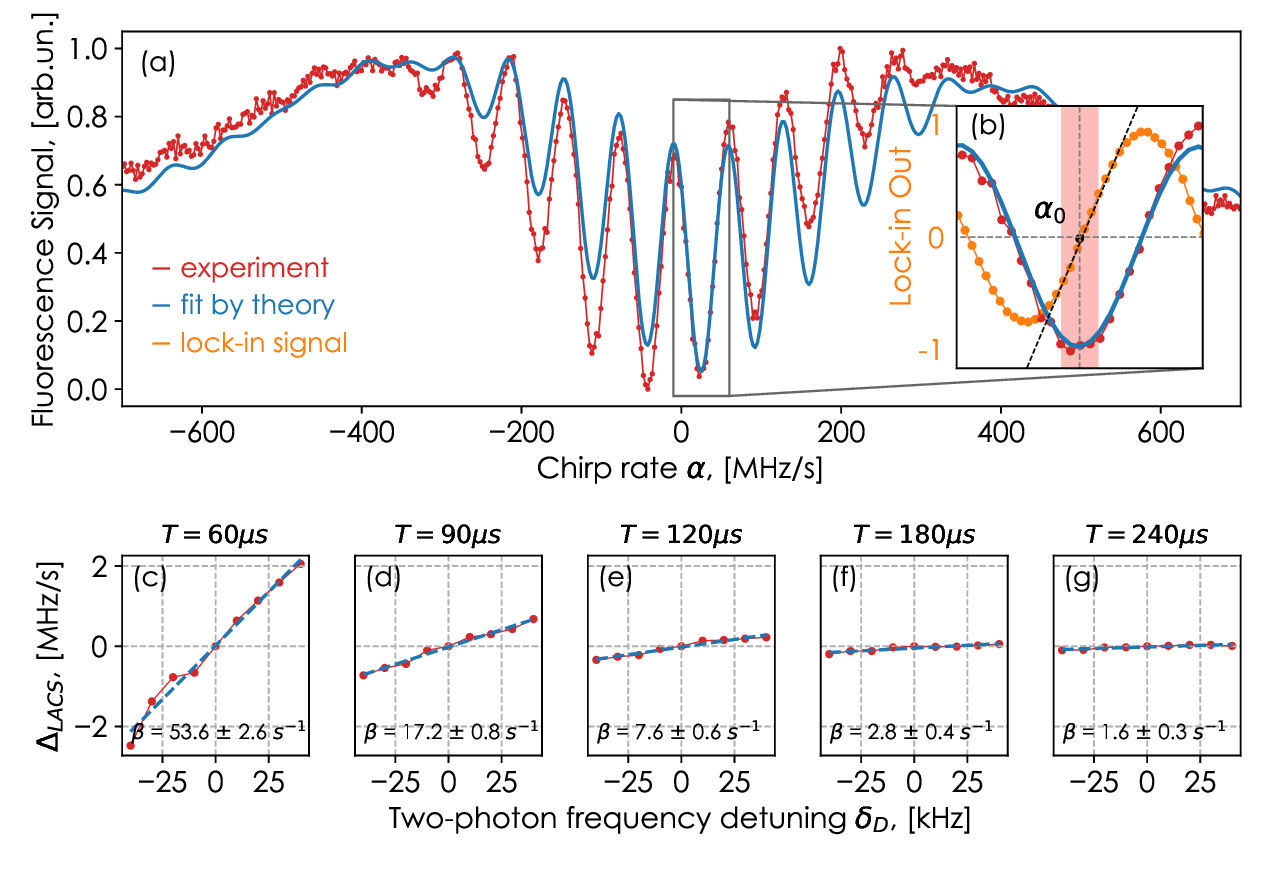}
    \caption{(Color online) Observation of the LACS shift in an atomic interferometer-gravimeter: (a) typical interference signal (red curve, blue curve — fit based on the theoretical model \cite{Yudin_2025}) obtained for $T=110~\mu$s without averaging (each point represents a single experimental cycle); (b) fragment of the central interference fringe (red curve, blue — fit) and the corresponding lock-in amplifier output signal (orange curve) fitted with a straight line (black dashed-dotted line) over its linear region (highlighted in pink); (c–g) plots of the LACS shift $\Delta_{LACS}$ versus Raman beam detuning $\delta_D$ for different free-evolution intervals $T$.}
    \label{fig:detection}
\end{figure*}

In the final stage, the atomic interferometer signal was detected. The atoms were sequentially irradiated with two probe pulses resonant with the $|F=2\rangle \leftrightarrow |F'=3\rangle$ transition. The pulse duration and the interval between them were 1~ms and 1.5~ms, respectively. The first pulse was used for the selective detection of atoms in the interferometer output state $|F=2\rangle$. The second pulse measured the total number of atoms. To achieve this, between the probe pulses, all atoms were transferred to the $|F=2\rangle$ state by a 100~$\mu$s repumping pulse resonant with the optical transition $|F=1\rangle \leftrightarrow |F'=2\rangle$. To suppress the effects of atom-number fluctuations and probe-light intensity instability, the interferometric signal measured during the first pulse was normalized to the total atom number measured by the second pulse. Fluorescence was recorded using a Touptek IUA1700KMA camera.

It should be noted that measurements of $g$ in atomic interferometer-gravimeters are performed by recording the interference signal during the frequency chirp of the Raman beams \cite{Peters_2001}. A typical interference signal is shown in Fig.~\ref{fig:detection}(a). The value of $g$ is calculated using the equation $g=2 \pi \alpha_0 / k_{eff}$, where $k_{eff}=|\bf{k_1}|+|\bf{k_2}|$ is the effective wave vector of the Raman beams with $\bf{k_1}$ and $\bf{k_2}$, and $\alpha_0$ is the chirp rate ($\approx 25.14$~MHz) corresponding to the minimum of the central interference fringe, whose position (in the absence of shifts) does not depend on the interrogation time $T$.

In our previous theoretical paper \cite{Yudin_2025}, we showed that the lineshape asymmetry leads to a shift of this minimum $\alpha_0$ (the LACS shift) and thus introduces a systematic error in the measurement of $g$. A demonstration of such asymmetry in the interference signal is shown in Fig.~\ref{fig:detection}(a). According to \cite{Yudin_2025}, this shift depends on the parameters of the atomic cloud, the Raman beams, and the interferometric sequence; in the case of a non-monochromatic atomic wave packet, it scales as:
\begin{equation}
\Delta_{LACS} \propto \frac{\pi}{2k_{eff}^2 \sigma_\upsilon^2} \frac{\Delta_D}{T^3}~,
\label{eq:shift}
\end{equation}
where $\sigma_\upsilon$ is the width of the atomic cloud velocity distribution and $\Delta_D$ is the residual Doppler detuning, equivalent to the detuning $\delta_D$ used in the present paper.

\begin{figure}
    \centering
    \includegraphics[width=0.95\linewidth]{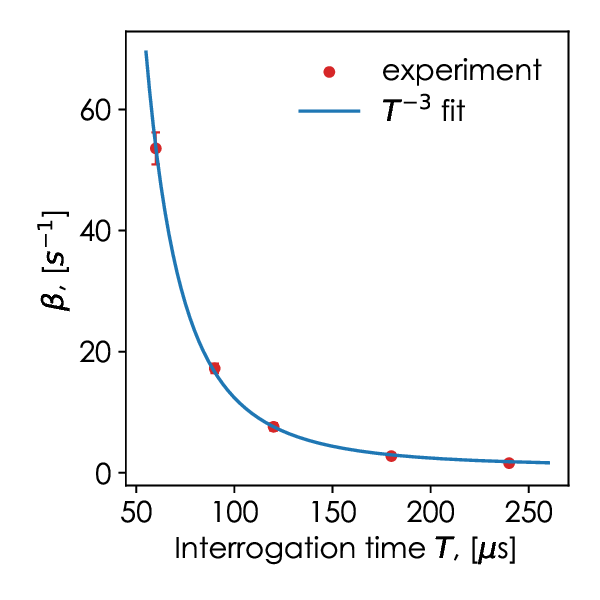}
    \caption{(Color online) Plot of the proportionality coefficient $\beta$ as a function of the free-evolution time $T$.}
    \label{fig:T3_shift}
\end{figure}

We have carried out an experimental study of the dependence of the shift $\Delta_{LACS}$ on the Raman beam detuning $\delta_D$ in an atomic interferometer for an initial atomic ensemble with a broad velocity distribution ($\sim$5~$\mu$K). The shift was defined as $\Delta_{LACS}(\delta_D) = \alpha_0(\delta_D) - \alpha_0(0)$, where $\alpha_0(\delta_D)$ is the the central interference fringe minimum position measured at the corresponding value of $\delta_D$. The detuning $\delta_D$ was varied within the range $\pm$40~kHz. The registration of the interference fringe minimum was performed using a lock-in detection method by scanning and modulating $\alpha$ near this minimum. A typical lock-in amplifier signal is shown in Fig.~\ref{fig:detection}(b). The scanning range corresponded to the interference signal period, equal to $1/T^2$~Hz/s. The modulation frequency and amplitude were approximately $\sim$1~Hz and $1/8T^2$~Hz/s, respectively. The value of $\alpha_0$ was measured by fitting the lock-in amplifier signal with a linear function over its linear region. Fig.~\ref{fig:detection}(c–g) show the measured $\Delta_{LACS}$ shift as a function of the frequency detuning $\delta_D$ for five different free-evolution intervals $T$: 60, 90, 120, 180, 240~$\mu$s. The selection of small $T$ values was motivated by the fact that the LACS shift follows an inverse cubic scaling $\propto 1/T^3$, while the atom interferometer sensitivity $\sigma_g\propto1/k_{eff}gT^2$ exhibits an inverse quadratic dependence. Thus, reducing $T$ leads to a linear increase in the relative LACS contribution compared to the measurement noise, which facilitates its experimental detection. The obtained dependencies were fitted with a linear function to determine the proportionality coefficient $\beta=\Delta_{LACS} / \delta_D$. Fig.~\ref{fig:T3_shift} presents the experimental $\beta(T)$ dependence, which is characterized by an inverse cubic behavior. For example, at $T=60$~$\mu$s, the measured coefficient was $\beta_{meas}=53.6\pm2.6$~s$^{-1}$, which agrees well with the calculated value $\beta_{calc}=49.1$~s$^{-1}$. Consequently, the presented results confirm that the observed $T^{-3}$-shift originates from LACS and are in good agreement with our theoretical calculations \cite{Yudin_2025}.

To the best of our knowledge, the accuracy requirements for compensating the Raman pulse frequency detuning $\delta_D$ have not been analyzed in the literature so far. However, as shown in this work, such compensation is essential for suppressing the associated LACS shift, especially in atomic interferometers with a short baseline. The minimization of this detuning is typically achieved either by monitoring the two-photon resonance signal, as performed in this work, or by maximizing the interference signal contrast. The estimated accuracy of these techniques is on the order of 10~kHz. As follows from our results, such an uncertainty can lead to a significant systematic errors in the measured value of $g$ due to the LACS effect. For example, for $T$ on the order of several milliseconds, typical for short-baseline atomic interferometer-gravimeters \cite{Lee_2022,McGuinness_2012}, the corresponding contribution may reach magnitudes of 0.1–1~mGal, which is comparable to their sensitivity and accuracy levels.

To date, various methods for suppressing systematic shifts have been successfully implemented in atomic gravimeters. For instance, the most widely used measurement protocol for $g$ \cite{Bidel_2013}, based on alternating the signs of the effective wave vectors $\bf{k_{eff}}$, enables compensation of first-order light shifts as well as shifts caused by magnetic field inhomogeneities. However, this method does not provide efficient cancellation of the LACS shift due to the potential asymmetry of the shifts for opposing $\bf{k_{eff}}$ vectors. Another commonly used approach, based on a single measurement of the shift followed by its subsequent correction, is also generally not optimal, since the magnitude of the LACS shift may vary over time due to drifts and fluctuations of the relevant parameters of the Raman beams and the atomic cloud. All of this motivates the development of effective methods for suppressing the LACS shift in atomic interferometers. For example, one possible approach is the use of hyper-Ramsey spectroscopy schemes \cite{Zanon_2022}, currently being developed by the co-authors of this work and potentially capable of compensating the LACS shift through modification of the spectroscopic sequence.

In conclusion, we have experimentally demonstrated for the first time the LACS shift in an atomic interferometer-gravimeter and confirmed its inverse cubic dependence on the atomic free evolution time. Our research shows that this effect can make a noticeable contribution to the systematic error of gravitational acceleration measurements in short-baseline atomic interferometers. The obtained results are consistent with our previously developed theoretical model and highlight the necessity of accounting for spectroscopic lineshape asymmetry when optimizing interferometer parameters, as well as the need for developing LACS shift suppression techniques. This paper provides an experimental basis for further studies of the considered $T^{-3}$-shift and its impact on the metrological performance of compact atomic interferometers.

\textbf{Acknowledgments.}
The authors thank S.~A.~Farnosov and V.~A.~Vasiliev for their assistance in the fabrication and adjustment of the electronic systems used in this study.



\end{document}